\documentclass[12pt,preprint]{aastex}

\slugcomment{Accepted for publication in ApJL}

\shorttitle{SPIFFI observations of SMM\,J14011+0252}
\shortauthors{Tecza et al.}

\begin{document}

\title{SPIFFI observations of the starburst SMM\,J14011+0252:\altaffilmark{1}\\
Already old, fat, and rich by z = 2.565}

\altaffiltext{1}{Based on observations obtained at the European Southern 
Observatory, Chile, for programme 70.A-0254.}

\author{M. Tecza\altaffilmark{2},
A. J. Baker\altaffilmark{2},
R. I. Davies\altaffilmark{2},
R. Genzel\altaffilmark{2,3},
M. D. Lehnert\altaffilmark{2}, \\
F. Eisenhauer\altaffilmark{2},
D. Lutz\altaffilmark{2},
N. Nesvadba\altaffilmark{2},
S. Seitz\altaffilmark{4},
L. J. Tacconi\altaffilmark{2}, \\
N. A. Thatte\altaffilmark{5},
R. Abuter\altaffilmark{2},
\& R. Bender\altaffilmark{2,4}
}

\altaffiltext{2}{Max-Planck-Institut f{\" u}r extraterrestrische Physik,
Postfach 1312, D-85741 Garching, Germany; \{tecza,ajb,davies,genzel, 
mlehnert,eisenhau,lutz,nicole,linda,rabuter,bender\}@mpe.mpg.de}

\altaffiltext{3}{Department of Physics, 366 Le~Conte Hall, University of 
California, Berkeley, CA 94720-7300}

\altaffiltext{4}{Munich University Observatory (USM), Scheinerstrasse 1,
D-81679 M{\" u}nchen, Germany; stella@usm.uni-muenchen.de}

\altaffiltext{5}{University of Oxford Astrophysics, Keble Road, Oxford OX1
3RH, United Kingdom; thatte@astro.ox.ac.uk}

\begin{abstract}
Using the SPectrometer for Infrared Faint Field Imaging (SPIFFI) on the ESO 
VLT, we have obtained $J$, $H$, and $K$ band integral field spectroscopy of 
the $z = 2.565$ luminous submillimeter galaxy SMM\,J14011+0252.  A global 
spectrum reveals the brighter of this spatially resolved system's two 
components as an intense starburst that is remarkably old, massive, and 
metal-rich for the early epoch at which it is observed.  We see a strong 
Balmer break implying a $\geq 100\,{\rm Myr}$ timescale for continuous star 
formation, as well as nebular emission line ratios implying a supersolar 
oxygen abundance on large spatial scales.  Overall, the system is rapidly 
converting a large baryonic mass into stars over the course of only a few 
hundred Myr.  Our study thus adds new arguments to the growing evidence that 
submillimeter galaxies are more massive than Lyman break galaxies, and more 
numerous at high redshift than predicted by current semi-analytic models of 
galaxy evolution.
\end{abstract}


\section{Introduction}\label{s-intro}

Deep imaging in the optical and near-IR has made it possible to constrain 
both the evolution of the cosmic star formation rate density (e.g., Madau et
al. 1996) and its time integral, the growth of the cosmic stellar mass density 
(e.g., Dickinson et al. 2003).  Short-wavelength studies give an incomplete 
picture of these trends, however, since an important population of 
high-redshift galaxies is too obscured to be easily detected.  These 
submillimeter galaxies (SMGs; see Blain et al. 2002 and references therein)
have large IR luminosities that are predominantly generated by star formation
rather than accretion \citep{barg01,alma03}; as a result, they contribute 
substantially to cosmic star formation \citep{barg00}.  Moreover, as the 
strikingly different appearances of the Hubble Deep Field at $0.83\,{\rm \mu 
m}$ \citep{will96} and $850\,{\rm \mu m}$ \citep{hugh98} exemplify, SMGs are 
rarer and forming stars much more intensely than typical optically selected 
systems.

To understand the evolutionary state of SMGs, we have conducted a detailed 
rest-frame optical case study using SPIFFI, the SPectrometer for Infrared 
Faint Field Imaging \citep{tecz00,eise00,eise03}.  Our target is 
SMM\,J14011+0252 (hereafter J14011), a $z = 2.565$ galaxy lying behind the 
$z = 0.25$ cluster A1835 and the second SMG to have an optical 
counterpart identification \citep{barg99} validated by CO interferometry 
\citep{fray99}.  Recent maps of J14011's molecular gas have led to 
divergent conclusions about its intrinsic size: \citet{ivis01} suggest that a 
large background source undergoes a factor ${\cal M} \sim 2.5$ magnification 
by A1835 as a whole, while \citet{down03} argue that a fortuitously located 
cluster member further boosts magnification of a small background source to 
${\cal M} \sim 25$.  We defer a full discussion of lensing to a future paper; 
here, we focus primarily on the lensing-independent conclusions of high age 
and metallicity that can be drawn from J14011's global spectrum alone.  
Throughout the paper we assume a flat $\Omega_\Lambda = 0.7$ cosmology with 
$H_0 = 70\,{\rm km\,s^{-1}\,Mpc^{-1}}$.

\section{Observations and data reduction}\label{s-obs}

In three SPIFFI ``Guest Instrument'' runs between February and April 2003, we 
observed J14011 in the $J$, $H$, and $K$ bands on Kueyen (UT2 of the ESO
VLT).  Weather conditions were extremely stable within and between nights, and 
the near-IR seeing was better than $0.6\arcsec$.  SPIFFI obtains simultaneous
spectra for all pixels in a $32 \times 32$ array; our data were obtained with 
the $0.25\arcsec$ pixel scale, and with spectral resolving power $R \equiv 
\lambda/\Delta \lambda \sim 1500$ at $1.3\,{\rm \mu m}$, $\sim 2000$ at 
$1.6\,{\rm \mu m}$, and $\sim 2400$ at $2.2\,{\rm \mu m}$.  Individual 
exposure times in the three bands were 10, 5, and 10 minutes, respectively; 
we obtained an ``off'' frame at a sky position (offset by $20\arcsec - 
25\arcsec$) for each ``on'' frame, and jittered the telescope by two-pixel 
offsets between the ``on'' frames themselves.  The total on-source 
integration times were 60 minutes in $J$, 95 minutes in $H$, and 340 minutes 
in $K$.

We reduced the data within IRAF using modified versions of 
long-slit spectroscopy tools.  After subtracting dark frames and
flat-fielding all data (object and sky frames), we masked out bad pixels and 
interpolated new values using all available three-dimensional information.  
Object and sky frames were then wavelength calibrated using arc lamp spectra
and/or night sky emission lines.  The ``off'' sky frames were then subtracted, 
and a linear fit to the background was subtracted across each spectral and 
spatial row of the data.  At this point the data cubes for each on-off pair 
were reconstructed, using only integer pixel shifts and assuming that each 
spatial row of the field covered exactly 32 pixels; although this was not 
strictly correct, it only slightly reduced the spatial resolution and avoided 
loss of signal-to-noise due to resampling.  A final background subtraction 
exploited the two-dimensional field of view: objects in the frame were masked 
out, and the mean of the remaining pixels in a given wavelength plane was 
subtracted from that plane.  Using telescope offsets, we spatially aligned and 
then combined the data cubes, clipping any deviant pixels in each cell.  
Telluric correction was applied to the final merged cubes.  Fluxes were 
calibrated by integrating SPIFFI observations of the UKIRT faint standard 
FS135 only over the Mauna Kea Observatory $J$, $H$, and $K$ filter bandpasses; 
the reference photometry of FS135 itself has a scatter of less than 1\% 
\citep{hawa01}.

\section{Results}\label{s-res}

\begin{figure}
\plottwo{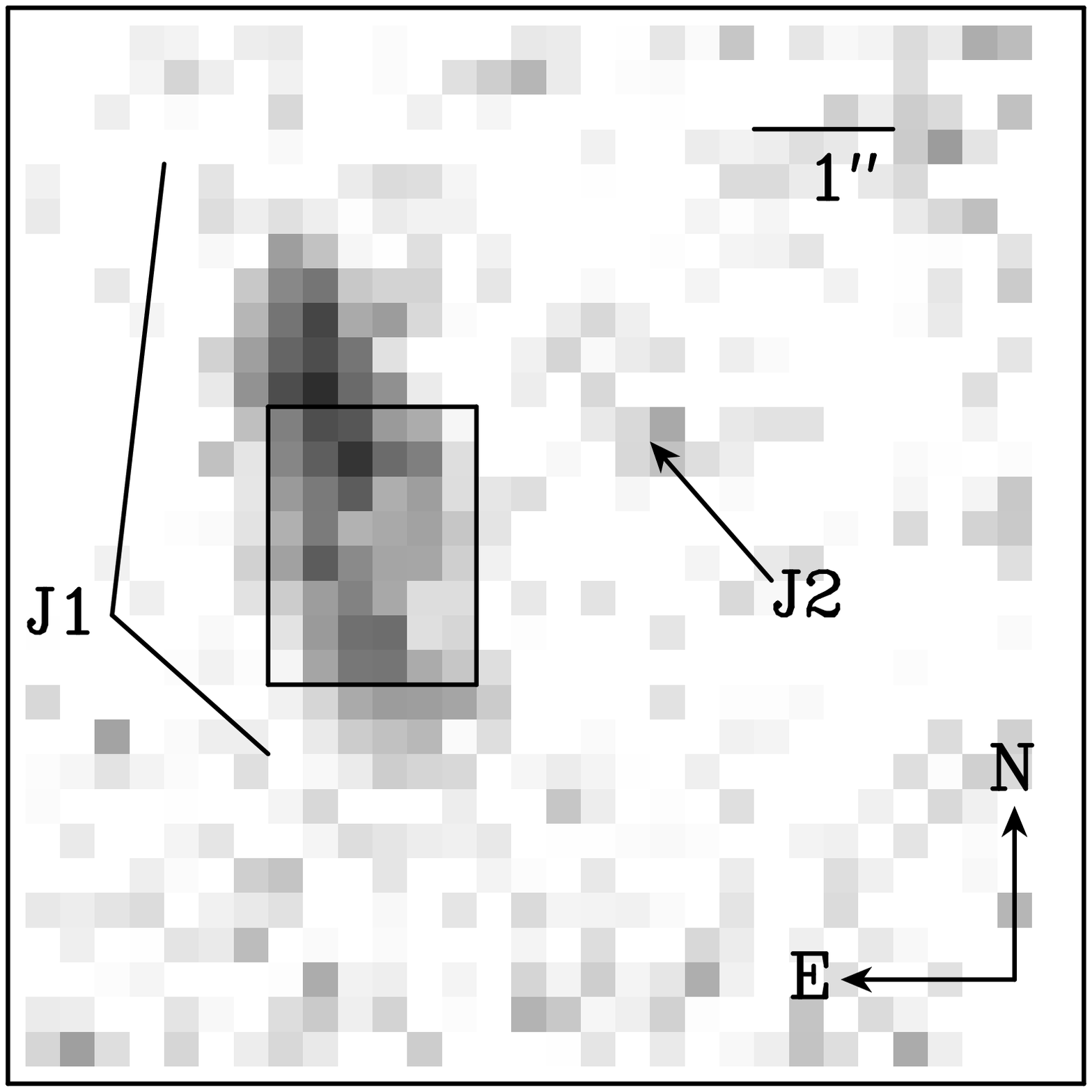}{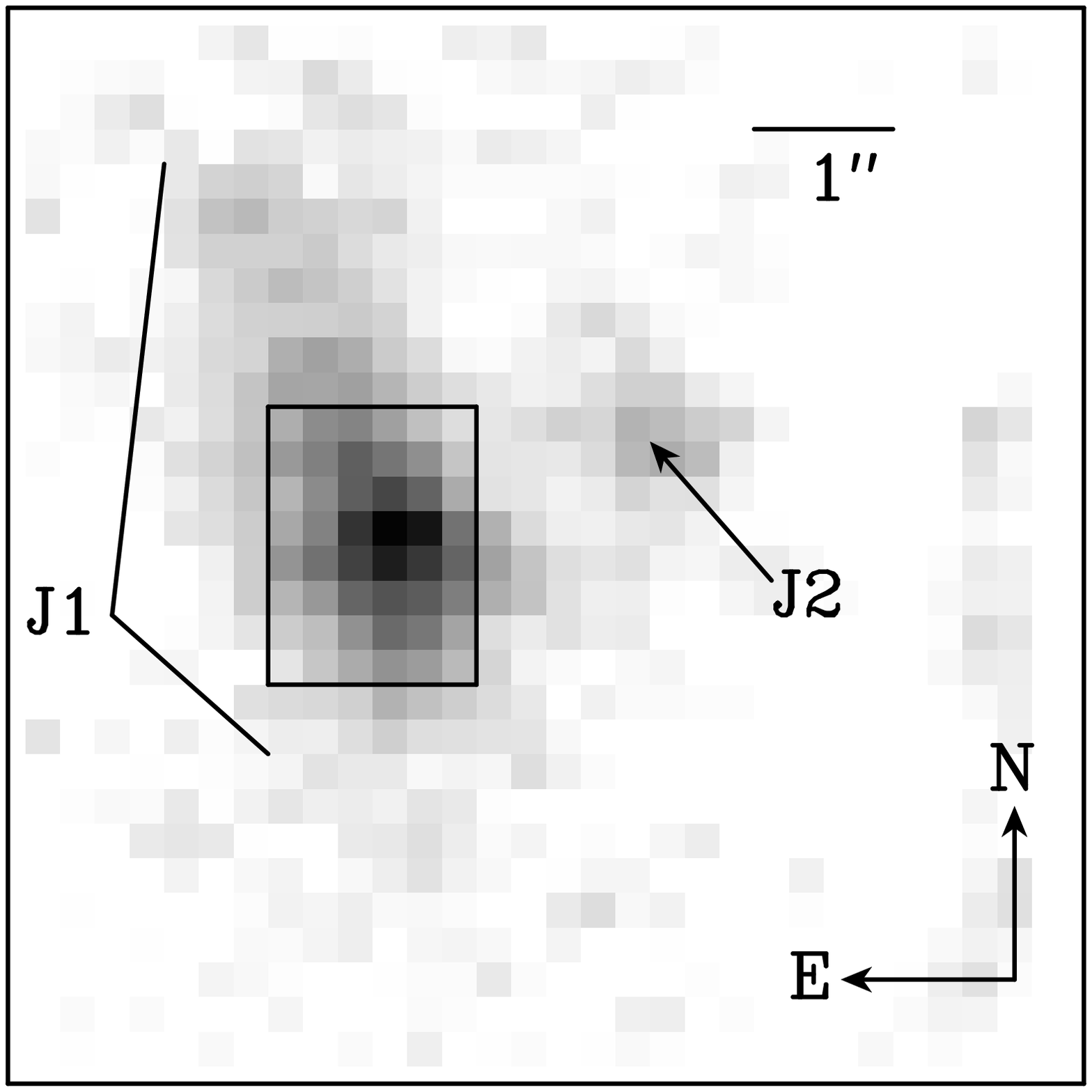}
\caption{Left: continuum-subtracted ${\rm H\alpha}$ image of J14011.  Right: 
line-free $K$-band continuum image of J14011.  The field of view is 
$8\arcsec \times 8\arcsec$; the aperture used to extract the composite 
spectrum for J1 in Figure \ref{f-jhkspec} is overlaid.
\label{f-maps}}
\end{figure}

\begin{figure}
\plotone{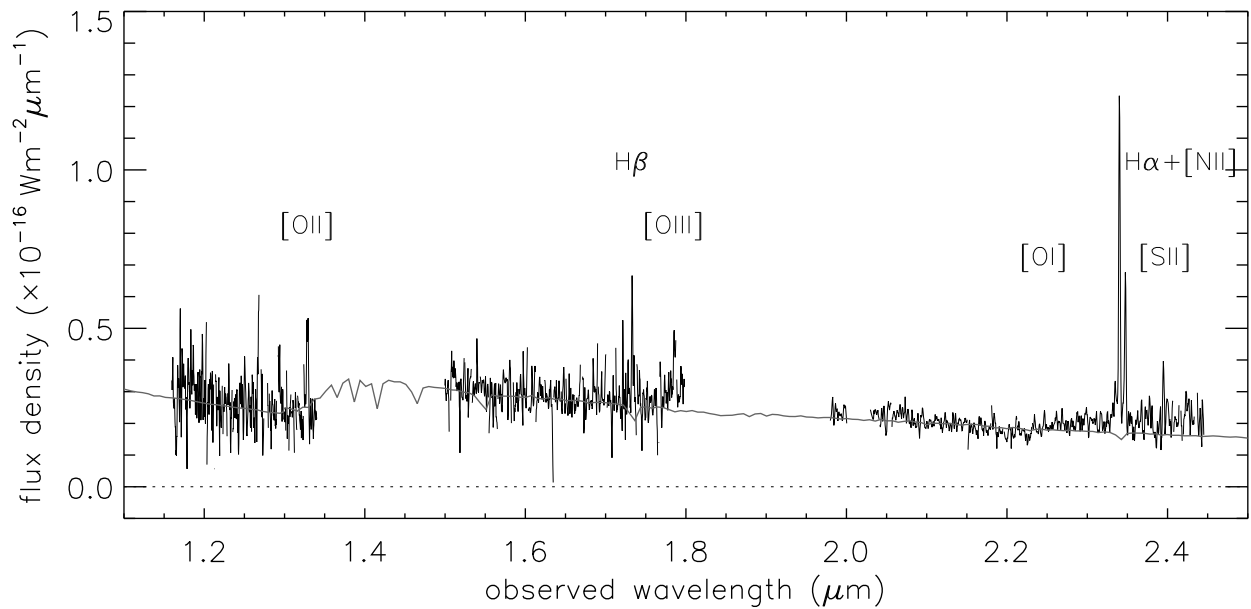}
\caption{$J$, $H$, and $K$ spectra of J14011 extracted from a $2\arcsec \times 
1.5\arcsec$ aperture.  Superposed is the model STARS spectrum for a 200\,Myr 
old continuous star formation episode, assuming solar metallicity, a 
$1-100\,M_\odot$ Salpeter IMF, and extinction $A_V = 0.7$.
\label{f-jhkspec}}
\end{figure}

Figure \ref{f-maps} shows J14011 in continuum-subtracted 
${\rm H\alpha}$ and line-free $K$-band continuum.  Within the
bright eastern ``J1'' component, the SPIFFI data reveal substantial spatial 
variation in ${\rm H\alpha}$ equivalent width: nebular emission is
strongest away from the central continuum peak.  After using this peak
to align all three data cubes, we extracted a spectrum of J1 
from the $2\arcsec \times 1.5\arcsec$ region shown in Figure \ref{f-maps}.  
This box encloses about two thirds of the total 
continuum associated with J1 in the $K$-band cube, and virtually 
all the emission we detect in $J$ and $H$.  Figure \ref{f-jhkspec} 
shows the result and reveals that J14011 has a continuum break between the
observed $J$ and $H$ bands.  The lack of a similar break in blank-sky 
spectra extracted from elsewhere in the SPIFFI field proves that this 
cannot be an artifact of poor sky subtraction; the $J-H$ color of FS135, 
meanwhile, is accurate to 0.3\% \citep{hawa01}.  The feature's 
most likely origin is the Balmer break of a $\geq 100\,{\rm Myr}$ 
old stellar population at $z \simeq 2.565$.  Quantitatively, the strength of 
the break can be reproduced by a population synthesis model constructed with 
the STARS package \citep{ster98} for a 200\,Myr old episode of 
continuous star formation with a $1-100\,M_\odot$ Salpeter initial mass 
function (IMF), solar metallicity, and an extinction $A_V = 0.7$.  A crucial 
implication of this result is that the near-IR continuum emission must be 
dominated by the high-redshift background source, rather than by a 
foreground cluster galaxy.

We have also used the spectrum in Figure \ref{f-jhkspec} to measure fluxes and 
centroids for eleven nebular emission lines; taken together, they 
imply J1 has a systemic redshift $z_{\rm H\,II} = 2.5652 \pm 0.0006$, in 
perfect agreement with the $z_{\rm CO}$ derived from millimeter interferometry 
\citep{down03}.  ${\rm H\alpha}$ emission from the fainter western ``J2'' 
component is blueshifted by $\Delta z = 0.0021 \pm 0.0003$ with respect to 
J1, consistent with the offset measured by \citet{ivis00} from rest-UV 
spectra.  The J1 line fluxes allow us to measure the $R_{23}$ estimator for 
its oxygen abundance \citep{page79}.  After 
correcting for reddening with a Galactic extinction curve \citep{howa83} and 
the color excess $E(B-V) = 0.18^{+0.13}_{-0.12}$ implied by the Balmer 
decrement, and fixing $I_{4959} \equiv I_{5007}/3$, we derive ${\rm 
log}(R_{23}) = 0.32 \pm 0.16$.  Translation of this index to an oxygen 
abundance depends on ionization parameter, as quantified by ${\rm log}(O_{32}) 
= -0.24 \pm 0.22$.  Using the fits of Kobulnicky, Kennicutt, \& Pizagno (1999) 
to the calibrations of \citet{mcga91}, we infer that $12 + {\rm log\,(O/H)} = 
8.96^{+0.06}_{-0.10}$.  Relative to the measurement of Allende~Prieto, 
Lambert, \& Asplund (2001), this represents an abundance that is supersolar by 
$0.27^{+0.11}_{-0.15}\,{\rm dex}$.  A high [\ion{N}{2}]/${\rm H\alpha}$ ratio 
means J14011 cannot fall on the lower-metallicity branch of the [O/H] 
vs. $R_{23}$ relation.

\section {The high mass of J14011} \label{ss-j1}

Although we reserve a full lensing model of J14011 for a future paper, our 
SPIFFI data already allow us to evaluate the small source/high ${\cal 
M}$ and large source/low ${\cal M}$ scenarios.  As noted 
above, the strength of the apparent Balmer break in J1 leaves little 
room for continuum emission from an ``extra'' lensing galaxy in A1835.  The 
newly confirmed velocity difference between J1 and J2, on top of ${\rm
  Ly}\,\alpha$ and C\,III] $1909\,{\rm \AA}$ equivalent
width differences in their published spectra \citep{barg99,ivis00}, indicates 
that these components are not multiple images of the same background source, 
contrary to the suggestion of \citet{down03}.  Finally, SPIFFI spatially 
resolves J1 in an east-west direction; the observed $0.95\arcsec$ 
FWHM of ${\rm H\alpha}$ emission exceeds the $0.65\arcsec$ east-west PSF 
dimension.  This would not occur if J1 were formed from the merging of 
a triple image set.

These direct arguments for a low lensing magnification are nicely complemented
by the wealth of self-consistent evidence that J14011 as a whole has a large 
mass and age.  Our estimate of J1's high oxygen abundance is 
independent of ${\cal M}$, provided there is minimal differential lensing 
from 3726 to $5007\,{\rm \AA}$ rest wavelength.  For a closed-box 
model with solar yield, this metallicity corresponds to 
a gas/baryonic mass fraction of ${\rm exp}\,(-Z/Z_\odot) \simeq 0.16$.  
\citet{fray99} measure a $8.2 \times 10^{10}\,{\cal M}^{-1}\,{\rm 
K\,km\,s^{-1}\,pc^2}$ CO(3--2) line luminosity for J14011.  For gas that is 
dense and warm 
enough to produce detectable CO(7--6) emission, \citet{down03} suggest that 
molecular gas mass and CO luminosity are related by $M_{\rm gas}/L^\prime_{\rm 
CO(3-2)} = 0.8\,M_\odot\,({\rm K\,km\,s^{-1}\,pc^2})^{-1}$.  Adopting ${\cal 
M} \sim 5$ as a fiducial value for the cluster-only lensing case, 
we infer a molecular gas mass $M_{\rm gas} \simeq 1.3 \times 
10^{10}\,M_\odot$, and therefore a total baryonic mass $M_{\rm 
bary} \simeq 7.9 \times 10^{10}\,M_\odot$ and stellar mass 
$M_{\rm stars} = M_{\rm bary} - M_{\rm gas} \simeq 6.6 \times 
10^{10}\,M_\odot$.  This last value (and with it, our assumption 
of moderate ${\cal M}$) is empirically supported by the mass-metallicity 
relation, as recently rederived by \citet{trem04} from a sample of $\sim 
51,000$ galaxies at $z \sim 0.1$.  Their favored translation of $R_{23}$ to 
metallicity gives $12 + {\rm log(O/H)} \simeq 9.1$ for J1, at which the 
median stellar mass is $\sim 6.2 \times 10^{10}\,M_\odot$.  

For a simple closed-box model, assumption of continuous star formation 
and a return fraction $R = 0.2$ uniquely sets the length of time for
which a galaxy has been forming stars: $t_{\rm SF} = \{{\rm 
exp}\,(Z/Z_\odot) - 1\}\,(1-R)^{-1}\,M_{\rm gas}\,{\rm SFR}^{-1}$.  For 
J14011, $t_{\rm SF}$ will be independent of lensing because 
$M_{\rm gas}$ and SFR depend identically on ${\cal M}$.  We can
estimate the star formation rate in J14011 from its $S_{850} = 12.3\,{\cal 
M}^{-1}\,{\rm mJy}$ \citep{smai02}; for the fiducial SED of \citet{blai99} 
with $T_{\rm d} = 40\,{\rm K}$, $\beta = 1.5$, and $\alpha = 2.2$, $L_{\rm IR} 
\simeq 1.9 \times 10^{12}\,(S_{850}/{\rm mJy})\,L_\odot = 2.3 
\times 10^{13}\,{\cal M}^{-1}\,L_\odot$.  Adjusting the 
conversion factor of \citet{kenn98} to match a $1-100\,M_\odot$ Salpeter
IMF, we infer ${\rm SFR} \simeq 1920\,{\cal M}^{-1}\,M_\odot\,{\rm yr}^{-1}$ 
and $t_{\rm SF} \simeq 220\,{\rm Myr}$.  This age is consistent 
with the strength of the Balmer break seen in Figure \ref{f-jhkspec}, 
suggesting that the closed-box assumption made above
is not unreasonable.  J14011's powerful starburst thus appears to be converting
most of its baryons into stars on a fairly rapid (few hundred Myr)
timescale.  

\section {The high masses of SMGs} \label{ss-bmat}

Its star formation history and metallicity allow J14011 to join 
SMM\,J02399-0136-- whose $M_{\rm bary}$ \citet{genz03} infer from its 
spatially resolved gas kinematics-- as an SMG whose 
large baryonic mass is secure.  Thanks to CO confirmation of Keck/LRIS 
redshifts obtained by \citet{chap03b}, this category promises to expand 
rapidly.  \citet{neri03} have published CO data confirming the Keck 
redshifts for three new SMGs; the total set of five has 
$\left<z\right> = 2.72$ and a median CO linewidth of $420\,{\rm km\,s^{-1}}$ 
(FWHM).  A fruitful comparison can be made between this SMG sample and the 
well-studied Lyman break galaxy (LBG) population at $z \sim 3$.  In the list 
of sixteen LBGs for which \citet{pett01} report ionized gas kinematics, the 
median velocity width is $170\,{\rm km\,s^{-1}}$ (FWHM) at a slightly larger 
$\left<z\right> = 3.12$.  Assuming that the (few kpc) size scales of the 
starbursts are comparable in LBGs and SMGs, the higher velocity widths imply 
SMGs' dynamical masses should also be larger by a typical factor of 6.  If 
the two populations' dark matter fractions are comparable, we would expect a 
median SMG baryonic mass exceeding the median $1.1 \times 10^{10}\,M_\odot$ 
stellar mass estimated for 74 LBGs \citep{shap01}\footnote{This median LBG 
mass adjusts the \citet{shap01} population synthesis results for a 
$1-125\,M_\odot$ Salpeter IMF and $H_0 = 70$.} by the same factor, i.e., $6.6 
\times 10^{10}\,M_\odot$.  Although the strong clustering of LBGs (e.g., 
Giavalisco \& Dickinson 2001) suggests they share their host halos with much 
more baryonic mass than is seen in their inner regions, SMGs appear already to 
have assembled measurably larger masses of baryons, or to have been more 
successful in retaining those baryons in the face of energetic winds.  
Complementary constraints on the star formation histories and metallicities of 
more SMGs, like those obtained here for J14011, would further strengthen this 
case.

In the hierarchical merging scenario for structure formation, massive 
galaxies are assembled in greater numbers with the passage of time.  As
several authors have pointed out, the observed number densities of 
high-redshift galaxies with large stellar masses exceed the predictions of
semi-analytic models of galaxy evolution within this framework 
\citep{cima02,dadd04,sara04}.  As in \citet{genz03}, we can also use the
number densities of SMGs to apply the same ``baryonic mass assembly test'' to
the models.  \citet{smai02} report the surface density of SMGs with 
$S_{850} > 5\,{\rm mJy}$ (the threshold above which SMG optical redshifts 
$1.8 \leq z \leq 3.5$ have been confirmed with CO interferometry) to be 
$880_{-330}^{+530}\,{\rm deg^{-2}}$.  The \citet{chap03b} sample includes two 
(of nine) sources above this threshold lying at $z < 1.8$, and excludes the 
$\sim 35 \pm 5\%$ of bright SMGs that lack radio counterparts 
\citep{ivis02,chap03a}.  Two of the five SMGs with CO detections have $\geq 
510\,{\rm km\,s^{-1}}$ FWHM line widths that would correspond to $M_{\rm bary} 
\geq 10^{11}\,M_\odot$ by the above rescaling of LBG stellar masses.
This leaves an estimated surface density of $180_{-70}^{+110}\,{\rm 
deg^{-2}}$ for SMGs with baryonic masses $\geq 10^{11}\,M_\odot$ in the range
$1.8 \leq z \leq 3.5$, and a comoving number density of $8.9_{-3.3}^{+5.3} 
\times 10^{-6}\,{\rm Mpc}^{-3}$.  

\begin{figure}
\plotone{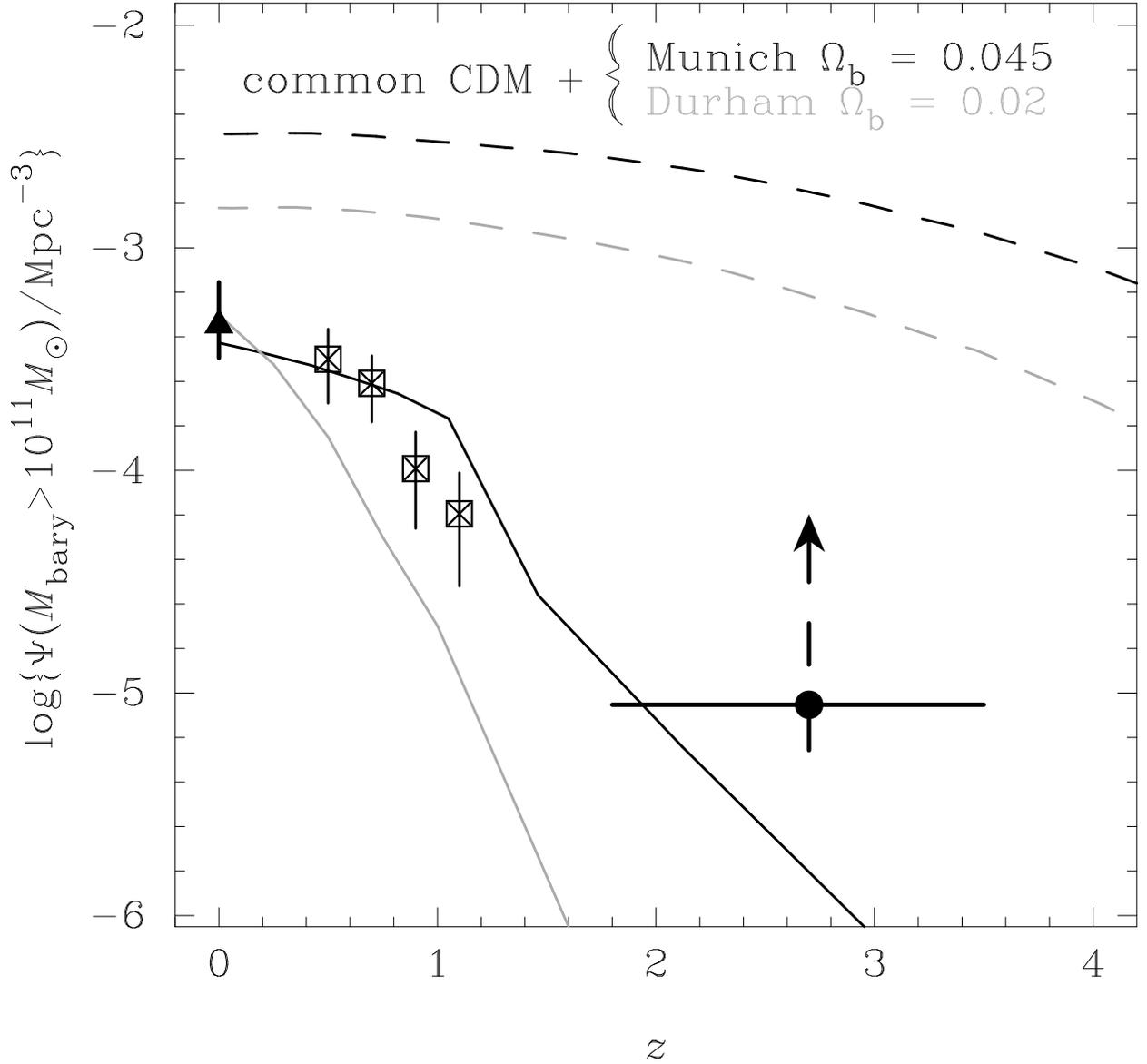}
\figcaption{Comoving number densities of galaxies with baryonic masses $\geq
10^{11}\,M_\odot$ as a function of redshift.  Triangle and open squares show 
densities of massive stellar systems at $z = 0$ \citep{cole01} and $z \sim 1$ 
\citep{dror04}; circle shows density for massive SMGs at $z \sim 2.7$, with a
factor of 7 correction for burst lifetime (see \S \ref{ss-bmat}).  Solid 
curves show the predictions of semi-analytic modelling by the ``Munich''
group \citep{kauf99} and the ``Durham'' group (an updated version of Baugh et
al. 2003); dashed curves show the corresponding number densities of halos with
{\it available} baryonic masses $\geq 10^{11}\,M_\sun$.  The two models use
the same halo simulations but assume different $\Omega_{\rm b}$.
\label{f-bmat}}
\end{figure}

We must still correct this density upward to account for sources that 
are comparably massive but no longer luminous enough to have been detected at 
$850\,{\rm \mu m}$.  Adding $t_{\rm SF}$ and the gas exhaustion timescale
$M_{\rm gas}\,{\rm SFR}^{-1}$ for J14011 implies that the total duration of
its current burst will be of order 250\,Myr.  Relative to the 1.8\,Gyr elapsed 
over the range $1.8 \leq z \leq 3.5$, this would imply a correction factor of 
7 for the population as a whole.  Figure \ref{f-bmat} plots this prediction, 
together with the comoving number densities of galaxies with stellar masses 
above the $10^{11}\,M_\odot$ threshold at $z \sim 0$ \citep{cole01} and $z 
\sim 1$ \citep{dror04}.  Also plotted are the predictions of semi-analytic 
models, one of which is taken from \citet{kauf99} and the other an update of 
\citet{baug03}.  Since all theoretical and observational values assume the 
same cosmology and a consistent (either Miller-Scalo or $1-100\,M_\odot$ 
Salpeter) IMF, we are secure in concluding that the models underpredict the 
number densities of massive galaxies at $z \sim 2.7$.  Couching Figure 
\ref{f-bmat} in terms of {\it mass} emphasizes that the observed surface 
densities of SMGs cannot be explained using a very flat IMF alone.  We also 
note that the models contain sufficiently many dark halos of total mass $\geq 
10^{11}\,(\Omega_{\rm M}/\Omega_{\rm b})\,M_\odot$ to account for the observed 
number densities of massive SMGs, provided their baryons can be rapidly and 
efficiently assembled into galaxies.

Correcting the observed number density of SMGs to account for
their less dust-luminous descendants raises the question of what types of 
galaxies SMGs can plausibly evolve into.  Although LBGs are clearly excluded 
due to their smaller masses, an intriguing alternative is the newly identified 
population whose red $J - K$ colors can stem from a strong Balmer break at 
$z \sim 2.5$ \citep{fran03}.  These galaxies appear to be strongly clustered 
\citep{dadd03} and-- assuming a uniform distribution in the redshift range 
$2 \leq z \leq 3.5$ \citep{vand03}-- have a comoving number density $\sim 1.8 
\times 10^{-4}\,{\rm Mpc^{-3}}$.  Relatively few sources with $J-K > 2.3$ 
also have 1.2\,mm flux densities $> 3\,{\rm mJy}$ (Dannerbauer et al. 2004),
indicating little direct overlap with the {\it bright} end of the SMG 
population (see also Frayer et al. 2004), although fainter (sub)millimeter 
counterparts may still be present (e.g., Wehner, Barger, \& Kneib 2002).  
With three examples in the HDF-S having (for a Miller-Scalo IMF) stellar 
masses $(0.6 - 1.4) \times 10^{11}\,M_\odot$ \citep{sara04}, it would seem 
plausible that the more luminous objects with red $J-K$ colors at this epoch 
could have passed through an SMG phase at higher redshift.  

\acknowledgments

We thank M. Horrobin, C. Iserlohe, A. Schegerer, and J. Schreiber for their 
hard work on SPIFFI; A. Sternberg for assistance with STARS; C. Baugh,
N. Drory, G. Kauffmann, and C. Tremonti for useful discussions; and an 
anonymous referee for helpful comments.

\clearpage

\end{document}